\documentclass[useAMS,usenatbib]{mn2e}
\usepackage{times}
\usepackage{hyperref}
\usepackage{graphicx}
\usepackage{amsmath}
\usepackage{amssymb}

\def\be{\begin{equation}}
\def\ee{\end{equation}}
\def\bea{\begin{eqnarray}}
\def\eea{\end{eqnarray}}

\title[Roche volume filling of star clusters in the Milky Way]{Roche volume filling of star clusters in the Milky Way}
\author[A. Ernst, A. Just]{A. Ernst$^{1}$\thanks{email: aernst@ari.uni-heidelberg.de}, A. Just$^{1}$ \\
$^{1}$Astronomisches Rechen-Institut am Zentrum f\"ur Astronomie der Universit\"at Heidelberg, 
M\"onchhofstrasse 12-14, 69120 Heidelberg, Germany}

\begin{document}

\date{Accepted ... Received ...}

\pagerange{\pageref{firstpage}--\pageref{lastpage}} \pubyear{2002}

\maketitle

\label{firstpage}

\begin{abstract}
We examine the ratios $r_h/r_J$ of projected half-mass and Jacobi radius 
as well as $r_t/r_J$ of tidal and Jacobi radius for open and globular clusters in the Milky Way using data of both observations and simulations. We applied an improved calculation of $r_J$ for eccentric orbits of globular clusters.
A sample of 236 open clusters of Piskunov et al. within the nearest kiloparsec around the Sun has been used.
For the Milky Way globular clusters, data are taken from the Harris catalogue.
We particularly use the subsample of 38 Milky Way globular clusters for which orbits have been integrated
by Dinescu et al. We aim to quantify the differences between open and globular clusters
and to understand, why they form two intrinsically distinct populations.
We find under certain assumptions, or, in other words, in certain approximations, 
(i) that globular clusters are presently Roche volume underfilling %, that the average open cluster
%is approximately {\bf Roche volume} filling, that a fraction of open clusters is {\bf Roche volume} overfilling 
and (ii) with at least $3\sigma$ confidence that the ratio
 $r_h/r_J$ of half-mass and Jacobi radius is $3 - 5$ times larger at 
present for an average open cluster in our sample than for an average globular cluster
in our sample and (iii) that a significant fraction of globular clusters may be 
Roche volume overfilling at pericentre with $r_t >
r_J$. Another aim of this paper is to throw light on the underlying theoretical reason for the existence of
the van den Bergh correlation between half-mass and galactocentric radius.
\end{abstract}


\raggedbottom

\section{Introduction}

Open star clusters (OCs) are abundant in the Milky Way disc. Their number is estimated to be of order $10^5$
\citep{Piskunov2006}. 
In contrast, there are approximately 150 globular clusters (GCs) known in the Milky Way \citep{Harris2010}.
While the GCs are orbiting on eccentric orbits with partly high inclinations with respect to the stellar disc 
plane of the Milky Way,
most OCs reside on near-circular orbits in the disc (although they may show a vertical oscillation with an amplitude of order
$\approx 0.5$ kpc \citep{Cararro1994}. The GCs are long-lived\footnote{If one considers them as ``living''.} with a 
mean age of 10 Gyr.
The OCs are short-lived with a mean lifetime of only 300 Myr \citep[][Figure 8.5, hereafter: BT2008]{Binney2008}. Moreover,
the lifetimes of OCs range from a few tens of Myr to a few Gyr. Age distributions are also given by \citet{Lamers2006}
and \citet{Bonatto2011}. 

\citet{Baumgardt2010} presented a weak evidence that there are two distinct GC populations outside the solar radius, namely a population of massive compact clusters with very small half-mass radii compared to the Jacobi radius and a population of low-mass and extended clusters with  $r_h/r_J > 0.1$. They argued that King models allow only a restricted range of 
$r_\mathrm{h}/ r_\mathrm{t}$ and used the half-mass radius to quantify the Roche volume filling of GCs. However, there are other dynamical models like polytropes \citep[see, e.g., ][]{Converse2010} with a much larger half mass radius compared to the tidal radius. Additionally Baumgardt et al. used the Jacobi radius of circular orbits instead of the more general definition provided by \citet{King1962} for eccentric orbits (see Appendix B4 for a more detailed comparison to the present work). 
We apply a more general derivation for circular and eccentric orbits including  $r_\mathrm{h}$ and $r_\mathrm{t}$ in order to distinguish the compactness of a cluster and the Roche volume filling factor.

It has been shown by \citet{vandenBergh1994}, that the half-light radii $r_h$ of GCs are correlated with the Galactocentric radius $R$
according to the relation

\be
r_h \propto R^{2/3} \label{eq:vdb}
\ee

\noindent
This relation has never been explained in terms of a deeper physical reason, although van den Bergh suspected already in
1994 that the correlation (\ref{eq:vdb}) could be imposed by the underlying galactic tidal field.
Assuming that the GCs are moving in an isothermal halo we find from analytical calculations that the time-dependent Jacobi radius 
(i.e., the distance to the Lagrange points $L_1$ and $L_2$) 
for eccentric orbits in such a halo scales as

\be
\lim_{R \gg L/V_C} r_{J, \, \rm isoth.} \propto R^{2/3} \label{eq:rjiso}
\ee

\noindent
(see appendix \ref{sec:iso}).
From the assumption of an isothermal halo, Eqns. (\ref{eq:vdb}) and (\ref{eq:rjiso}) and 
$\langle R \rangle \gg \langle L\rangle/V_C$, where $\langle R \rangle,  \langle L\rangle$ and $V_C$ are median Galactocentric radius, orbital angular momentum and the halo's circular velocity, respectively,
it is possible to conclude that GCs in the Milky Way are characterized by a ratio $r_h/r_J$ which is independent 
of $R$. The consequence is that, within the scatter of the correlation (which may be due to the scatter in GC masses), 
GCs are characterized by a common average relative size.
We note that the Jacobi radius (i.e., the distance from the cluster centre to the
Lagrange points $L_1$ and $L_2$) provides a natural scale
for star clusters in the tidal field. Other (dependent) scales are given by the positions/widths of the dominant 
resonances in the star cluster.

To gain a better understanding of the difference between OCs and GCs and why they form two distinct populations,
one may ask: Which values takes ratio $r_h/r_J$ on in the case of OCs.
While for GCs on eccentric orbits we may assume that their size is enforced by quantities in the pericenter, it seems reasonable to
suspect that OCs on near-circular orbits are not strongly influenced by the orbital evolution.
 
In the following discussions, we denote the ratio $r_{h}/r_J$ of 3D half-mass radius and Jacobi radius 
with the letter $\lambda$. We also define $\widehat{\lambda} = r_t/r_J$ 
using the cluster cutoff radii $r_t$ (i.e. the radii at which the density drops to zero) 
instead of the projected half-mass radii $r_h$.

The 3D half-mass radius $r_h=r_{h,\rm 3D}$ of a spherically symmetric stellar system 
is typically larger than the projected (2D) half-mass radius $r_{h,\rm 2D}$.
%(similar to the ``effective radius'' $R_e$ in the S\'ersic law) 
We have $r_{h,\rm 3D} \approx 1.3 \, r_{h,\rm 2D}$. For the Plummer model $r_{h, \rm 3D}/r_{h,\rm 2D} \approx 1.30$ 
can be obtained analytically.

This paper is organized as follows: Section 2 presents the theory, section 3 lists the current status of observations
and simulations. In Section 4 the results are calculated and Section 5 contains the discussion and conclusions.

\section{Theory}

Following the approach by \citet{King1962}, the Jacobi radius of a star cluster in the tidal field of a galaxy 
(i.e. the distance from the star cluster center to the Lagrange points $L_1$ and $L_2$) can generally be written as

\be
r_J = \left[ \frac{G M_{cl}}{\Omega^2 - \frac{d^2\Phi}{dR^2}} \right]^{1/3} \label{eq:geniso2}
\ee

\noindent
where $G$, $M_{\rm cl}$, $\Omega$, $\Phi$ and $R$ are the gravitational constant, the star cluster mass,
the angular speed, the gravitational potential of the galaxy and the galactocentric radius, respectively.
The last closed (critical) equipotential surface through the Lagrange points $L_1$ and $L_2$ encloses
the Roche or Hill volume.

Furthermore we define the ``(circular) velocity radius''
 
\be
r_v  = \frac{GM_{cl}}{V_{\rm C}^2} \label{eq:rv}
\ee

\noindent
It is the length scale at which the Keplerian circular velocity in the cluster (assuming a point mass cluster potential) 
would be equal to the circular velocity $V_{\rm C}$ in the Milky Way. 

The orbital periods $t_{\rm orb}$ and $T_{\rm orb}$ of a star at the half-mass radius of the star cluster and that of 
the star cluster orbit around the galaxy, respectively, are given by

\bea
t_{\rm orb}&=&\pi t_{\rm cr} = \frac{2\pi r_h}{\sigma_0} \approx 2\pi \sqrt{\frac{2 r_h^3}{GM_{\rm cl}}} \label{eq:kepler}
\ \ \ \mathrm{and}  \label{eq:torb1} \\
T_{\rm orb}&=&\frac{2\pi}{\Omega_C} = \frac{2\pi R_C}{V_C},  \label{eq:torb2}
\eea

\noindent
where $t_{\rm cr}$ is the crossing time (i.e. the time needed for a star to cross the half-mass sphere) 
and $\sigma_0^2\approx GM_{\rm cl}/(2 r_h)$ in dynamical equilibrium.

\subsection{Circular orbits}

For an OC on a circular orbit with radius $R_{\rm C}$ and velocity $V_C$ in a tidal field the Jacobi radius from 
Eqn. (\ref{eq:geniso2})
can be written as \citep{King1962, Kuepper2008, Just2009}

\be
r_J = \left[\frac{GM_{\rm cl}}{(4-\beta_C^2)\Omega_C^2}\right]^{1/3}
= \left(4-\beta_C^2\right)^{-1/3} r_v^{1/3} R_{\rm C}^{2/3}  \label{eq:rj1}
\ee

\noindent
with $\beta_C=\kappa_C/\Omega_C$ (see appendix \ref{sec:betac}), where $G$, $M_{\rm cl}$,  $\kappa_C$, $\Omega_C$ and $R_C$ are the gravitational constant, the total mass of the star cluster, the epicyclic and the circular frequency and the obital radius,
respectively, and $r_v$ is given by Eqn. (\ref{eq:rv}).

\noindent
Inserting (\ref{eq:torb1}) and (\ref{eq:torb2}) into (\ref{eq:rj1}) yields

\be
\lambda_{\rm OC}=\left(\frac{r_{h}}{r_J} \right)_{\rm OC} =  \left(\frac{4-\beta_C^2}{2}\right)^{1/3} \left(\frac{t_{\rm orb}}{T_{\rm orb}}\right)_{\rm OC}^{2/3} \label{eq:ratio1}
\ee

\noindent
for star clusters on circular orbits.

\begin{figure}
\includegraphics[angle=90,width=0.5\textwidth]{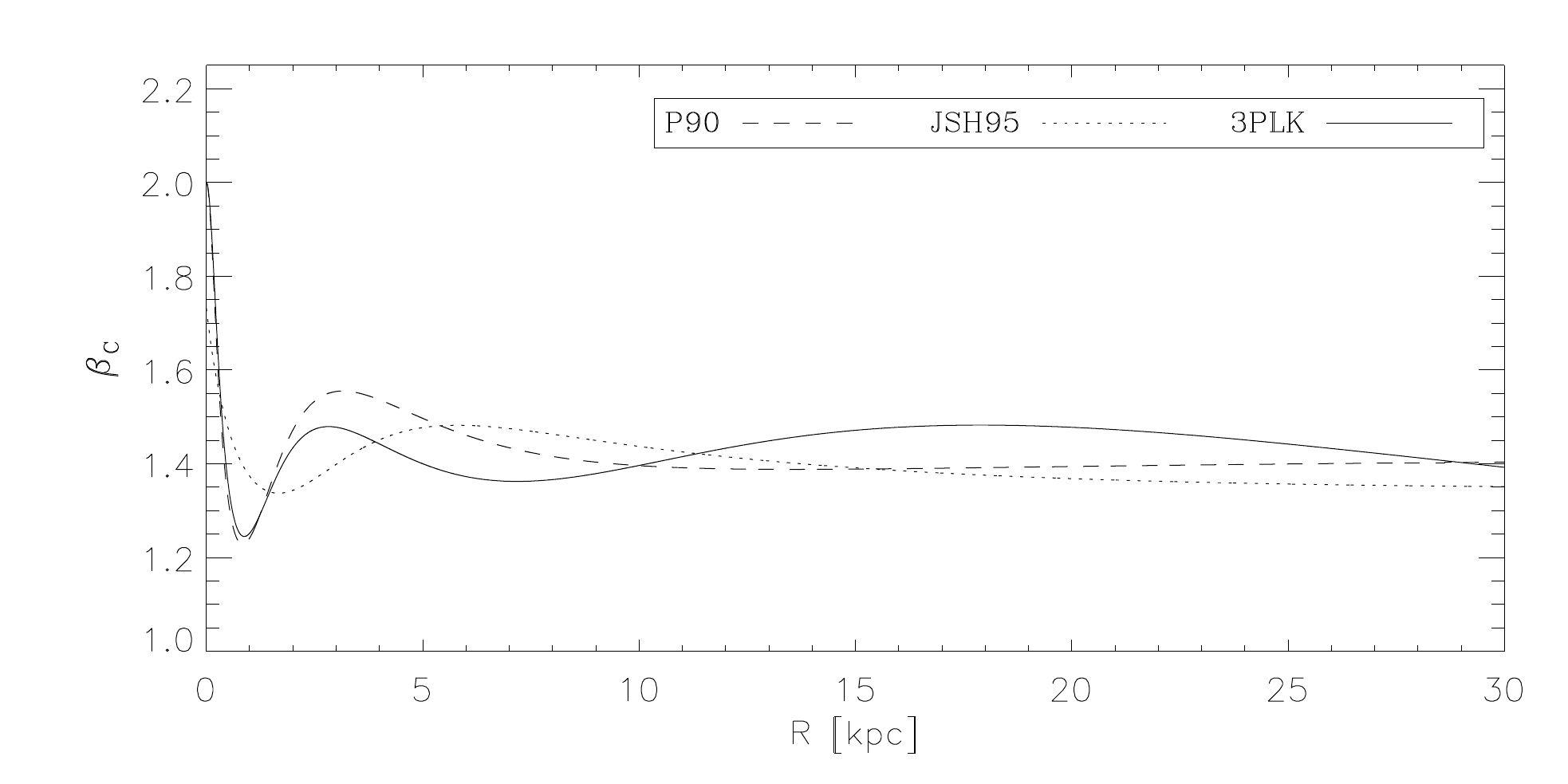} 
\caption{The ratio $\beta_C$ at $z=0$ for the JSH95 and P90 models of \citet{Dinescu1999} and the 
three-component Plummer-Kuzmin (3PLK) model used in \citet{Kharchenko2009} and \citet{Just2009}.} 
\label{fig:betaall}
\end{figure}

While $r_J$ determines the geometry of equipotential 
surfaces and cannot be observed, $r_h$, $t_{\rm cr} \propto r_h/\sigma_0$ and $T_{\rm orb}$ can be determined by observations and simulations.
For realistic Milky Way models, the ratio $\beta_C$ is approximately
 constant as a function of Galactocentric radius and the dependency on it is so weak (see Figure \ref{fig:betaall} and appendix \ref{sec:betac}) 
 that we can safely neglect its variation beyond $2-3$ kpc of the center of the Milky Way.

\bigskip\bigskip
\subsection{Eccentric orbits}

For GCs on eccentric orbits in an isothermal halo ($V_C=\mathrm{const}$) we must extend the theory (see appendix \ref{sec:iso}). 
Following the approach by \citet{King1962} the Jacobi radius from Eqn. (\ref{eq:geniso2}) can be written as

\be
r_J = r_v^{1/3} \left(\frac{R^4}{R_g^2 + R^2} \right)^{1/3} 
\ee

\noindent
where $R=R(t)$ is time-dependent, $r_v$ is given by Eqn. (\ref{eq:rv}), $V_C$ is the rotation speed of the isothermal sphere, $R_g=L/V_C$ is a guiding radius and the angular momentum $L=L(V_C,R_P,R_A)$ 
is a constant of motion given by Eqn. (\ref{eq:angmom}).
From Eqns. (\ref{eq:geniso2}) and (\ref{eq:kepler}) we also find

\bea
\lambda_{\rm GC} &=& \left( \frac{r_{h}}{r_J} \right)_{\rm GC} = \left(\frac{t_{\rm orb}}{2\pi}\right)^{2/3} \left[\frac{L^2}{R(t)^4} + \frac{V_C^2}{R(t)^2} \right]^{1/3}  \label{eq:ratio2} \\
&=& \left(\frac{t_{\rm orb}}{2\pi}\right)^{2/3}\left(\frac{V_C^2}{2}\right)^{1/3} \left(\frac{R_g^2 + R(t)^2}{R(t)^4}\right)^{1/3}.
\eea 

\begin{figure}
\includegraphics[width=0.5\textwidth]{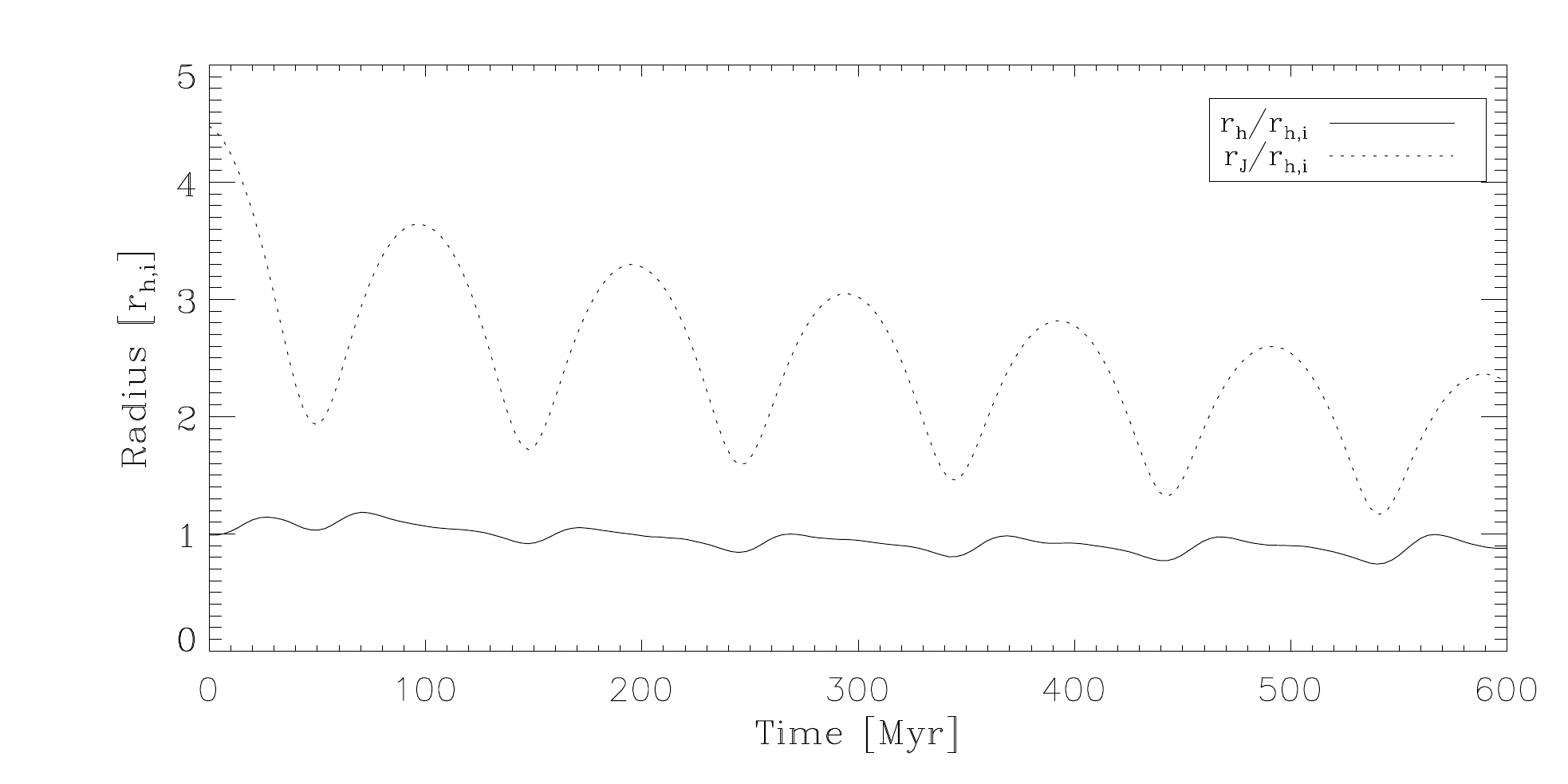} 
\caption{Time evolution of half-mass and Jacobi radii for a direct $N$-body simulation of a 
star cluster with $N=50000$ particles on an eccentric orbit within the disc plane. The cluster has a 
standard double-segment 
\citet{Kroupa2001} IMF and is initially Roche volume filling, i.e. the cutoff radius equals the Jacobi radius. 
The half-mass radius is calculated with respect to the current mass within the initial Jacobi radius. 
Both radii are scaled with the initial half-mass radius
$r_{h,i}\approx 9$ pc.} 
\label{fig:radii}
\end{figure}

We distinguish between two cases:

\begin{enumerate}
\item With the van-den-Bergh correlation (\ref{eq:vdb}) we obtain from Eqn. (\ref{eq:kepler})
along one orbit

\be
t_{\rm orb} \propto R \label{eq:vdbprop}
\ee

\noindent
and 

\be
\frac{t_{\rm orb}(t)}{R(t)} = \frac{t_{\rm orb,P}}{R_P} = \frac{t_{\rm orb, A}}{R_A}, \ \ \ \ \ \frac{t_{\rm orb}}{T_{\rm orb}} = \mathrm{const}
\ee

\noindent
where the subscripts ``P'' and ``A'' stand for peri- and apocenter. From Eqn. (\ref{eq:ratio2}) we obtain with (\ref{eq:vdbprop})

\be
\lambda_{\rm GC}^3 %\propto \frac{AR^2+ BR^4}{CR^4}
\propto \frac{A}{R^2} + B
\ee

\noindent
where $A$ and $B$ are two constants.
It follows that

\be
\lim_{R\gg R_g}\lambda_{\rm GC} = \mathrm{const}, \ \ \ \ \ \lim_{R \ll R_g}\lambda_{\rm GC} \propto R^{-2}
\ee

\noindent
where $R_g$ is given by Eqn. (\ref{eq:guiding}).

\item Contrariwise, direct $N$-body simulations suggest that for a bound GC the half-mass radius changes more slowly than the time scale $T_{\rm orb}$ while the Jacobi radius oscillates on that time scale according to Eqn. (\ref{eq:gen}) (see Figure \ref{fig:radii}).
This would mean that, if $M_{cl}\approx \mathrm{const}$,

\be
t_{\rm orb} \approx \mathrm{const} \label{eq:constprop}
\ee

\noindent
in difference to Eqn. (\ref{eq:vdbprop}).
 From Eqn. (\ref{eq:ratio2}) we obtain with (\ref{eq:constprop})

\be
\lambda_{\rm GC}^3 %\propto \frac{AR^2+ BR^4}{CR^4}
 \propto \frac{C}{R^4} + \frac{D}{R^2}
\ee

\noindent
where $C$ and $D$ are two constants.
\end{enumerate}

\section{Observations and simulations}

\label{sec:simobs}

\subsection{Samples and medians for OCs and GCs}

\begin{table}
\caption{Values for OCs and GCs. The values are medians if not denoted otherwise. 
The errors are standard errors of the median, i.e. divided by $\sqrt{N}$
of the sample size except for the error  $\Delta T_{\rm orb,OC}$. The data for OCs are derived 
from \citet{Piskunov2007}
and BT2008. The data for GCs are derived 
from \citet{Harris2010} and \citet{Dinescu1999}.}
\label{tab:clust-par}
\begin{center}
%\hspace{-1.5cm}
\begin{tabular}{ll}
\hline
OC parameter & Value \\
\hline
Sample size $N_{\rm OCs}$ & 236 \\
%Mean half-mass radius $r_h$ [pc] &  $2.59 \pm 0.15$ \\
Median projected half-mass radius $r_{h, \rm 2D}$ [pc] &  $1.94 \pm 0.15$ \\
Median tidal radius $r_t$ [pc] &  $7.90 \pm 0.51$ \\
Velocity dispersion $\sigma_0$ [pc Myr$^{-1}$] & $0.31$ \\ 
%Median relative error $\Delta t_{\rm cr, OC}/t_{\rm cr, OC}$ & $0.08$ \\
%Mean crossing time $t_{\rm cr, OC} $ [Myr] & $8.35 \pm 0.50$ \\
Median crossing time $t_{\rm cr, OC} = 2 r_{h, \rm 2D}/\sigma_0$ [Myr] & $12.52 \pm 0.96$ \\
%Relative error $\Delta T_{\rm orb, OC}/T_{\rm orb, OC}$ & $0.14$ \\
Average orbital period $T_{\rm orb, OC}$  [Myr] & $220 \pm 30$ \\
Average eccentricity $e_{\rm OC}$ & $0.127 \pm 0.003$ \\ 
\hline
GC parameter & Value \\
\hline
Sample size $N_{\rm GCs}$ & 34 (38) \\ %(157) \\
%Mean half-light radius $r_h$ [pc] & $4.38 \pm 0.54$ ($4.00 \pm 0.31$) \\
Median half-light radius $r_{h,\rm 2D}$ [pc] & $3.13\pm 0.51$ \\ % $(2.78 \pm 0.30)$ \\
Median tidal radius $r_t$ [pc] & $33.02 \pm 4.83$ \\ % $(2.78 \pm 0.30)$ \\
%Mean velocity disp. $\sigma_0$ [pc Myr$^{-1}$] & $10.43 \pm 1.23$ ($6.44 \pm 0.72$)  \\
Median velocity disp. $\sigma_0$ [pc Myr$^{-1}$] & $5.11 \pm 0.64$ \\ % ($5.37 \pm 0.64$)  \\
%Relative error of the mean $\Delta t_{\rm cr, GC}/t_{\rm cr, GC}$ & $0.29$  $(0.13)$ \\
%Median relative error $\Delta t_{\rm cr, GC}/t_{\rm cr, GC}$ & $0.62$ \\ % $(0.16)$ \\
%Mean crossing time $t_{\rm cr, GC}$ [Myr] & $0.70 \pm 0.20$ ($0.62 \pm 0.08$) \\
Median crossing time $t_{\rm cr, GC} =  2 r_{h, \rm 2D}/\sigma_0$ [Myr] & $1.175 \pm 0.726$ \\ %$(0.45 \pm 0.07)$ \\
%Relative error of the mean$\Delta T_{\rm orb, GC}/T_{\rm orb, GC}$ & $0.31$ \\
%Median relative error $\Delta T_{\rm orb, GC}/T_{\rm orb, GC}$ & $0.17$ \\
%Mean orbital period $T_{\rm orb, GC}$ [Myr] & $570 \pm 178$ \\
Median Galactocentric radius $R_{\rm orb, GC}$ [kpc] & $ 7.75 \pm 0.84$ \\
Median height above the disc plane $z_{\rm GC}$ [kpc] & $4.00$ \\ 
Median velocity $V_{\rm GC}$ [pc Myr$^{-1}$] & $175 \pm 16$ \\
Median orbital period $T_{\rm orb, GC}$ [Myr] & $207 \pm 54$ \\
%Median angular speed $\Omega_{\rm GCs}$ [Myr$^{-1}$] & $0.0259 \pm 0.0036$ \\ %= V_{\rm GC}/R_{\rm GC}
%Mean eccentricity $e_{\rm GC}$ & $0.551 \pm 0.028$ \\
Median eccentricity $e_{\rm GC}$ & $0.622 \pm 0.044$ \\
\hline
\end{tabular}
\end{center}
\end{table}

 For OCs we use the sample of 236 OCs of \citet{Piskunov2007}. We obtain 
 the values given in the upper part of Table \ref{tab:clust-par}.
 The projected half-mass radii of the 236 OCs have been obtained by solving 236 transcendental  equations
 with the Newton-Raphson method using the table of core and tidal radii provided by \citet{Piskunov2007} at the CDS.
The transcendental equation is given by 
 
 \be
 \ln X_h + \frac{X_h}{C} - 4\sqrt{\frac{X_h}{C}} -\frac{\ln C}{2} - \frac{1}{2C}  - \frac{2}{\sqrt{C}} + \frac{3}{2} = 0 \label{eq:transz}
 \ee
 
 \noindent
 with 
 
 \be
 X_h = 1 + (r_h/r_c)^2 \ \ \ \mathrm{and} \ \ \ C = 1 + (r_t/r_c)^2
 \ee
 
 \noindent
where $r_c, r_h$ and $r_t$ are the core, half-mass and tidal cutoff radii, respectively \citep[cf.][Section 3]{Ernst2010}.
Eqn. (\ref{eq:transz}) is solved for $r_h$.

The error of the median value of the projected half-mass radius $r_{h,\rm 2D}$ is defined as
 
 \be
 \Delta r_{h, \rm 2D} = \sqrt{\frac{\sum_{i=1}^{N_{\rm OCs}} \left[r_{h,i, \rm 2D} - \mathrm{Median}(r_{h,i, \rm 2D})\right]^2}{N_{\rm OCs}(N_{\rm OCs}-1)}} \label{eq:errmed}
 \ee

\noindent
We do not have an error on the velocity dispersion $\sigma_0$ of OCs. 
Thus we have set $\Delta\sigma_0 = 0$ for OCs.
Note also that in a 
bound system $r_h$ and $\sigma_0$ are related through the virial theorem, i.e. we have
$\Delta\sigma_0/\sigma_0=(1/2) \Delta r_h/r_h$ in virial equilibrium. We neglect this correlation since the virial theorem is not generally valid for star clusters in a tidal field. Namely, it is not valid for Roche volume overfilling star clusters.
The derived crossing time and its relative error are given by
\be
t_{\rm cr} = 2 r_{h,\rm 2D}/\sigma_0, \ \ \ \Delta t_{\rm cr}/t_{\rm cr} = \sqrt{\left(\Delta r_h/r_h\right)^2 + \left(\Delta \sigma_0/\sigma_0\right)^2} \label{eq:errder}
\ee

\noindent
The OCs in the sample by \citet{Piskunov2007} have approximately the orbital period of the Local
Standard of Rest (LSR) taken from BT2008, table 1.2. 

For GCs we use the median values from 34 out of a sample of 38 Milky Way globular clusters from 
\citet{Dinescu1999}.  The 
data compilation of 157 Milky Way GCs by Harris
(\citet{Harris2010}) was also used. 
The median values are obtained from the equivalents of Eqns. (\ref{eq:errmed}) and (\ref{eq:errder})
for the observed/simulated and the derived quantities. 
We obtained the values given in the bottom part of Table \ref{tab:clust-par}. Only for 97 out of 157 GCs from the Harris catalogue 
$\sigma_0$ is given. For 5 of these 97 GCs $r_h$ is not given as well.
From the GC sample in \citet{Dinescu1999} with 38 GCs we further find a median value
of the orbital period of GCs. Only for 34 out of 38 GCs in the Dinescu et al. sample $\sigma_0$ is given in the Harris catalogue.
The minimum Galactocentric radius is $R = 2.7$ kpc for NGC 6144 such that we can neglect the dependency on $\beta_C$
according to Figure \ref{fig:betaall}.  We further remark that a few recently discovered GCs have been added to
Harris' compilation by \citet{Ortolani2012}.

%For the further statistical analysis we define the ``effective sample size''

%\be
%N_{\rm eff} = \frac{N_{\rm OCs}N_{\rm GCs}}{N_{\rm OCs} + N_{\rm GCs}}
%\ee

%\noindent
%We have $N_{\rm eff} \approx 29.72$.

\section{Results}

\subsection{OCs}

\label{sec:ocs}

\begin{figure}
\includegraphics[width=0.5\textwidth]{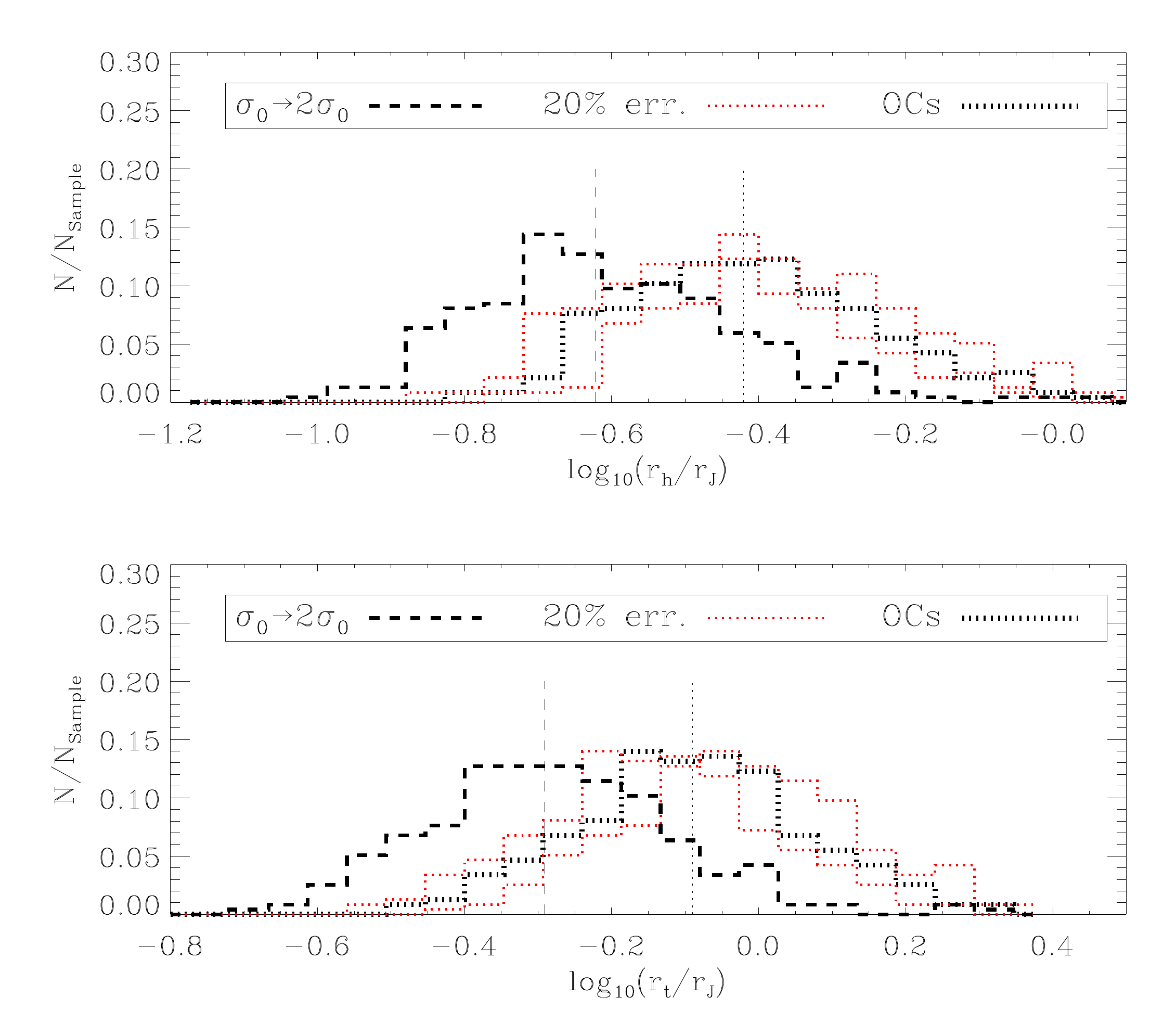} 
\caption{The ratios $r_{h,\rm  3D}/r_J$ (top panel) and $r_t/r_J$ (bottom panel) for OCs calculated from Eqn. (\ref{eq:ratio1})
for the sample of \citet{Piskunov2007}. The histograms for a canonical $\pm 20$\% error on $\sigma_0$
and for the substitution $\sigma_0\rightarrow2\sigma_0$ in Eqn.
(\ref{eq:ratio1}) are also shown. The vertical lines denote the medians. Note that $r_{h, \rm 2D}$ of the data set 
is corrected with a factor 
1.3 to obtain $r_{h, \rm 3D}$.}  
\label{fig:histmain}
\end{figure}

\begin{table}
\caption{Quantiles for OCs.}
\label{tab:OC-par}
\begin{center}
%\hspace{-1.5cm}
\begin{tabular}{llll}
\hline
Parameter & $10^{Q_{10}}$ & $10^{Q_{50}}$ & $10^{Q_{90}}$ \\
\hline
$\lambda_{\rm OC}$, Table \ref{tab:clust-par} &  0.239  &    0.379   &    0.675    \\
$\lambda_{\rm OC}$, $\sigma_0\rightarrow 2\sigma_0$ & 0.151 &    0.239 &     0.425   \\
$\widehat{\lambda}_{\rm OC}$, Table \ref{tab:clust-par} &  0.506 &     0.812 &      1.35 \\
$\widehat{\lambda}_{\rm OC}$, $\sigma_0\rightarrow 2\sigma_0$ &   0.319 &   0.511  &    0.852 \\
\hline
\end{tabular}
\end{center}
\end{table}

Figure \ref{fig:histmain} shows the distributions of $\log_{10}\lambda_{\rm OC}$ (top panel) and $\log_{10}\widehat{\lambda}_{\rm OC}$ (bottom panel) for OCs according to Eqn. (\ref{eq:ratio1}). We used the isothermal approximation $\beta_C=\sqrt{2}$. 
There is a large scatter of the sizes around the medians. We find also Roche volume overfilling
OCs.

$\widehat{\lambda}_{\rm OC}$ was found from Eq. (\ref{eq:ratio1}) with the values of Table \ref{tab:clust-par} using the tidal radii instead of
the half-mass radii in the definitions of crossing time (i.e, $t'_{\rm cr} = 2 r_t/\sigma_0$ instead of 
$t_{\rm cr} = 2 r_h/\sigma_0$). 

Figure \ref{fig:histmain} shows also the distributions of $\log_{10}\lambda_{\rm OC}$ with the substitution 
$\sigma_0\rightarrow2\sigma_0$ which leads to $t_{\rm cr, OCs}\rightarrow t_{\rm cr, OCs}/2$
in Eqn. (\ref{eq:ratio2}).

We find the medians $Q_{50}$ given in Table \ref{tab:OC-par}.
The quantiles $Q_{10}, Q_{50}$ and $Q_{90}$ have been calculated with an {\sc idl} routine by Hong \citep{Hong2004}.

%\begin{itemize}
%\item From Table  \ref{tab:clust-par}: $\lambda_{\rm OC} \approx 0.31$, $\widehat{\lambda}_{\rm OC} \approx 0.81$
%\item With $\sigma_0\rightarrow2\sigma_0$: $\lambda_{\rm OC} \approx 0.20$, $\widehat{\lambda}_{\rm OC} \approx 0.51$
%\end{itemize}

\subsection{GCs}

\label{sec:gcs}

\begin{figure}
\includegraphics[width=0.5\textwidth]{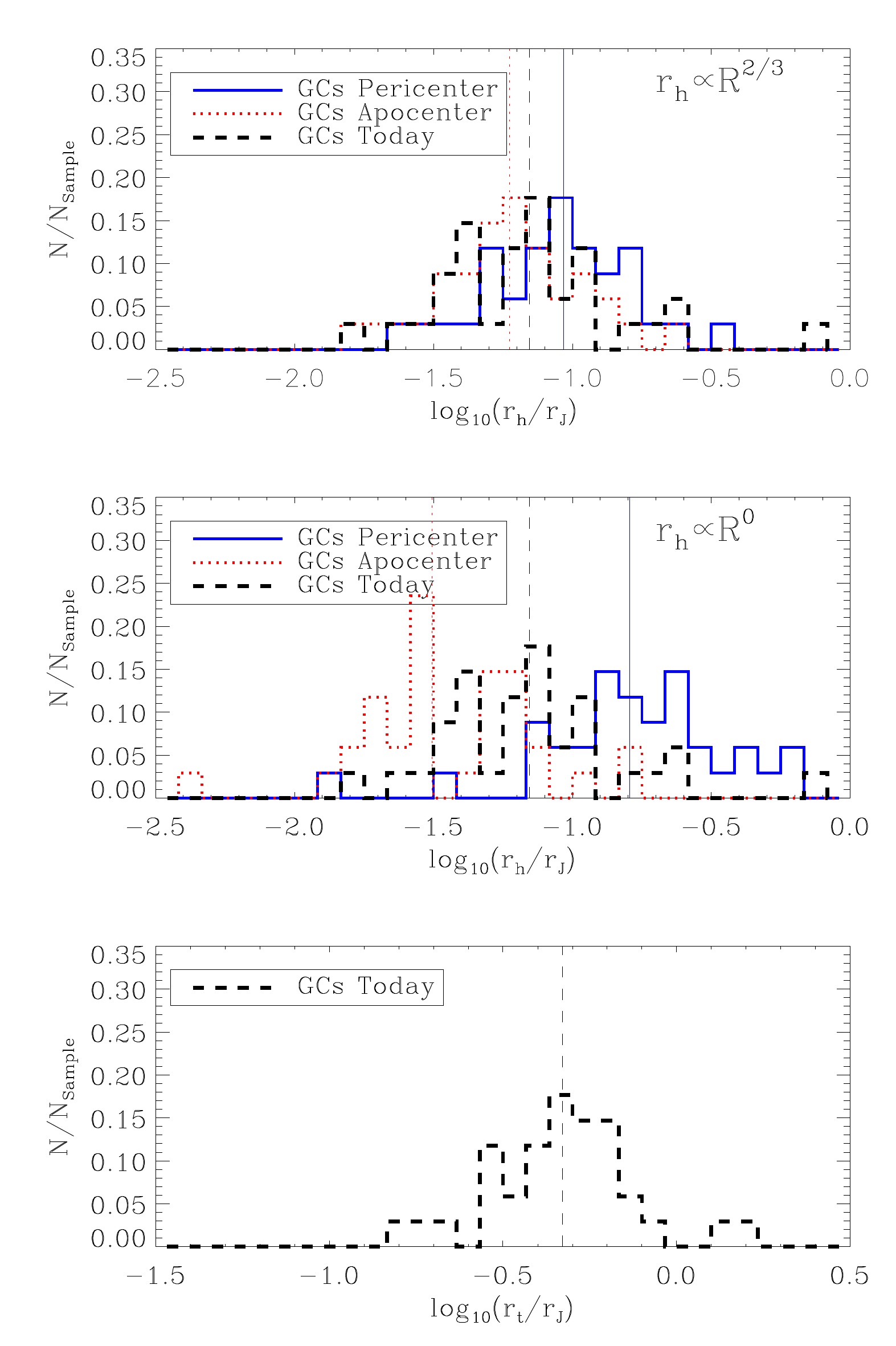} 
\caption{Top panel: The ratio $r_{h, \rm 3D}/r_J$ of GCs today and in the peri- and apocenter  calculated from Eqn. (\ref{eq:ratio2})
 for 34 GCs out of the sample by \citet{Dinescu1999}
under the assumption that the GCs are moving in an isothermal halo with $V_C\approx 228.5$ km/s and that the van-den-Bergh correlation holds. Middle panel: The same as in the top panel assuming that instead of the van-den-Bergh correlation 
$r_h$ is independent of Galactocentric radius. 
Bottom panel: The ratio $r_t/r_J$ for GCs today calculated from Eqn. (\ref{eq:gen}). The vertical lines denote the medians.
Note that $r_{h, \rm 2D}$ of the data set is corrected with a factor 1.3 to obtain $r_{h, \rm 3D}$.} 
\label{fig:histsgcs3}
\end{figure}

\begin{table}
\caption{Quantiles for GCs. The subscript ``0'' denotes present-day values.}
\label{tab:GC-par}
\begin{center}
%\hspace{-1.5cm}
\begin{tabular}{lclll}
\hline
Parameter & & $10^{Q_{10}}$ & $10^{Q_{50}}$ & $10^{Q_{90}}$ \\
\hline
$\lambda_{\rm GC,0}$ &  & 0.0318 &   0.0701 &     0.239  \\
$\widehat{\lambda}_{\rm GC,0}$ &  & 0.286   &  0.469 &   0.742  \\
\hline
$\lambda_{\rm GC,P}$ & ($r_h\propto R^{2/3}$) &   0.0427 &    0.0928  &    0.178 \\
$\lambda_{\rm GC,A}$ & " &  0.0297 &    0.0593 &    0.125  \\
\hline
$\lambda_{\rm GC,P}$ &($r_h\propto R^{0}$) &  0.0785   &     0.161  &    0.455 \\
$\lambda_{\rm GC,A}$ & " &  0.0174   &    0.0313   &    0.0847  \\
\hline
\end{tabular}
\end{center}
\end{table}

The top and middle panels of Figure \ref{fig:histsgcs3} show the distributions of $\log_{10}\lambda_{\rm GC}$ today and in the peri- and apocenter  calculated from Eqn. (\ref{eq:ratio2})  for 34 GCs of the sample by \citet{Dinescu1999}
under the assumptions that the GCs are moving in an isothermal halo with $V_C\approx 228.5$ km/s. In the top panel, we
assumed that the van-den-Bergh correlation is valid, and for the middle panel we assumed that $r_h\propto R^0$, i.e.
that it is independent of $R$.

%If the average GC can be fitted by a King model with $W_0=6$ 
%and $r_h/r_t=0.15$ according to Table 1 in G\"urkan, Freitag \& Rasio 2004
%we obtain the result that GCs are on average {\bf Roche volume} filling in the pericenter with $r_t=r_J$. 

The bottom panel of Figure \ref{fig:histsgcs3} shows the distribution of $\log_{10}\widehat{\lambda}_{\rm GC}$ today calculated from Eqn. (\ref{eq:gen})  for the same sample using the $r_t$'s given in the Harris catalogue.
It seems as if the GCs are today Roche volume underfillling and, moreover, at most Roche volume filling with 
$\widehat{\lambda}_{\rm GC} = 0.3 - 1.0$.

We find the medians $Q_{50}$ given in Table \ref{tab:GC-par}.

\subsection{Comparison of OCs and GCs}

\begin{figure}
\includegraphics[width=0.5\textwidth]{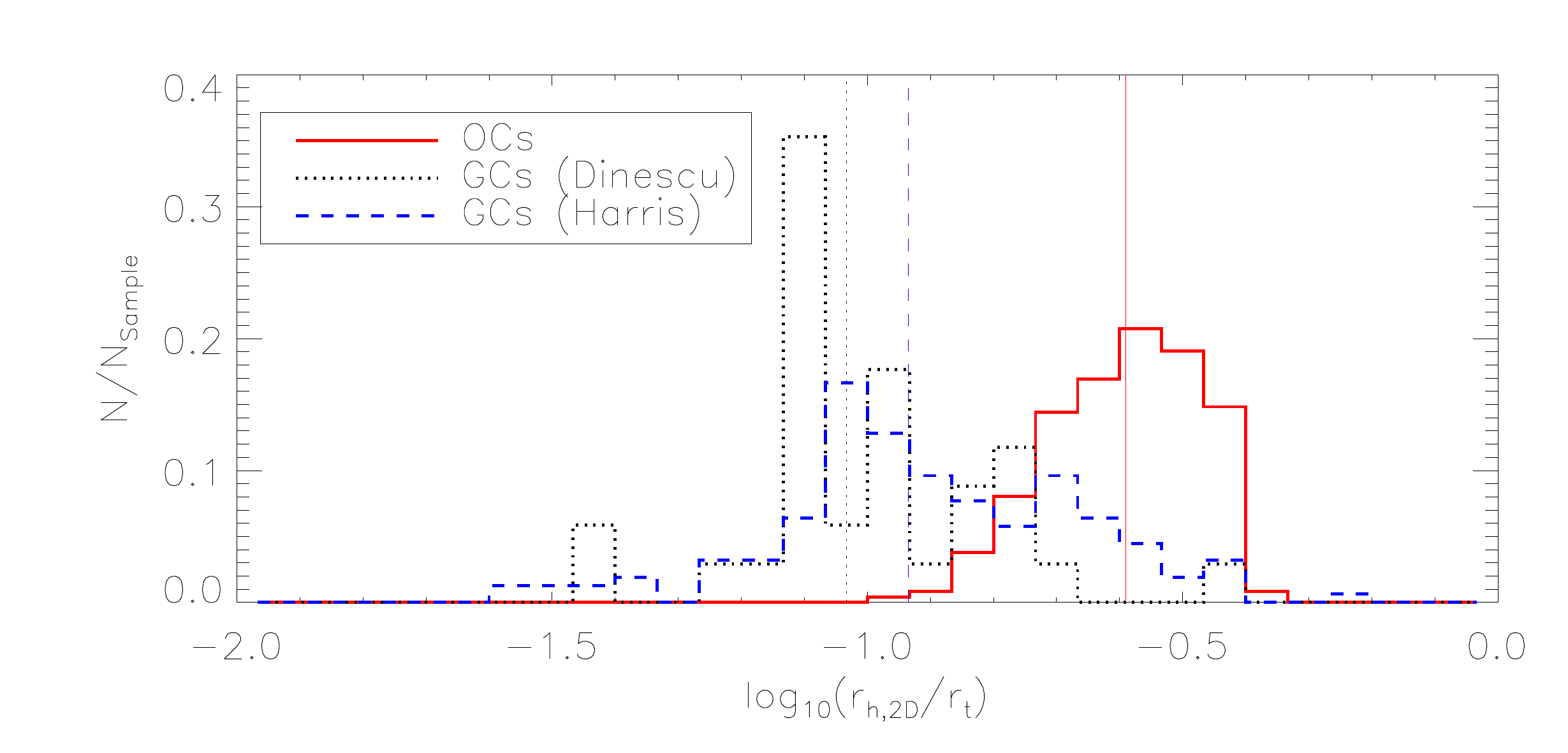} 
\caption{Top panel: The ratio $r_{h, \rm 2D}/r_t$ for OCs and GCs today 
 for the 236 OCs of the sample by \citet{Piskunov2007}, the 34 GCs of the sample by \citet{Dinescu1999}
 and 156 GCs of the Harris catalogue \citep{Harris2010}.
Note that $r_h=r_{\rm h, 2D}$ is in this Figure the projected half-mass radius (OCs) or the half-light radius (GCs) given in the 
Harris catalogue, respectively. The vertical lines denote the medians.} 
\label{fig:histrtrh}
\end{figure}

Figure \ref{fig:histrtrh} shows the ratio $r_{h, \rm 2D}/r_t$ for OCs and GCs today 
 for the 236 OCs of the sample by \citet{Piskunov2007}, the 34 GCs of the sample by \citet{Dinescu1999}
 and 156 GCs of the Harris catalogue \citep{Harris2010}.
Note that $r_h=r_{\rm h, 2D}$ is the projected half-mass radius (OCs) or the half-light radius (GCs) given in the 
Harris catalogue \citep{Harris2010}, respectively. The vertical lines denote the medians.
%The difference between OCs and GCs seems to be that, with respect to the concentration, the average OC has
%low concentration while the average GC has a higher concentration. 

We find the medians $\langle r_t/r_h\rangle = \langle\widehat{\lambda}/\lambda \rangle\approx 3.9$ (OCs),
$\langle r_t/r_h\rangle \approx 11.0$ (Dinescu GCs), $\langle r_t/r_h\rangle \approx 8.7$ (Harris GCs). 
Therefore, with respect to this ratio, the average OC corresponds to 
a King model with low $W_0$ while the average GC corresponds to a King model with high $W_0$ 
according to Table 1 in \citet{Guerkan2004}. 
We remark that the projected (2D) half-mass radius $r_h=r_{\rm h, 2D}$  is larger than the
3D half-mass radius $r_{\rm h, 3D}$. 
For the Plummer model $r_{\rm h, 3D}/r_{\rm h, 2D} \approx 1.30$ can be obtained analytically. 

Applying Eqns. (\ref{eq:ratio1}) and (\ref{eq:ratio2}) to the data %which we analyzed in Section \ref{sec:simobs} 
we find that

\be
\lambda_{\rm GC} < \lambda_{\rm OC}, \ \ \ \widehat{\lambda}_{\rm GC} < \widehat{\lambda}_{\rm OC}. \label{eq:tend}
\ee

\noindent
for an ``average''\footnote{``Average'' means here that its parameters are identical with the parameter 
median values of the whole sample.} 
GC and an ``average'' OC, with proportionality factors

\be
\mu = \frac{\lambda_{\rm OC}}{\lambda_{\rm GC}}, \ \ \ \widehat{\mu} = \frac{\widehat{\lambda}_{\rm OC}}{\widehat{\lambda}_{\rm GC}}  
\ee

\noindent
with $\mu \approx 5.4$ today and $\widehat{\mu}\approx 1.7$ today
from the median values of crossing and orbital times given in Tables \ref{tab:clust-par}, \ref{tab:OC-par}
and \ref{tab:GC-par}. 
The isothermal approximation $\beta_C=\sqrt{2}$ was used.

The relative error in $\mu$ is given by

\bea
\frac{\Delta\mu}{\mu}&=& 
\sqrt{\left(\frac{\Delta\lambda_{\rm OC}}{\lambda_{\rm OC}}\right)^2 + \left(\frac{\Delta\lambda_{\rm GC}}{\lambda_{\rm GC} }\right)^2}  \\
&=& \ln(10)\sqrt{ \left[\frac{  \Delta\log_{10}\lambda_{\rm OC}  }{\sqrt{N_{\rm OCs}}} \right]^2 +  \left[\frac{  \Delta\log_{10}\lambda_{\rm GC}   }{\sqrt{N_{\rm GCs}}} \right]^2 } \nonumber \\
&\lesssim& \ln(10)\sqrt{ \left[\frac{ (Q_{90} - Q_{10})_{\rm OCs}  }{\sqrt{N_{\rm OCs}}} \right]^2 +  \left[\frac{ (Q_{90} - Q_{10})_{\rm GCs}   }{\sqrt{N_{\rm GCs}}} \right]^2 } \nonumber
\eea

\noindent
The last line containing the 10 and 90 percent quantiles of the $\log_{10}\lambda$-distributions serves as an upper limit to the error. 
%We neglected the fact that the $\lambda$-distributions may be asymmetric due to the intrinsic scale $r_J$.
We obtain $\Delta\mu/\mu\lesssim 0.23$ for the upper limit.

%The confidence level $\nu \sigma$ is given by

%\be
%\nu = \frac{\mu-1}{\mu}\cdot \left(\frac{\Delta\mu}{\mu}\right)^{-1}.
%\ee

We further remark that taking $t_{\rm cr, OC}/2$ instead of $t_{\rm cr, OC}$ in Eqn. (\ref{eq:ratio1}) with all other quantities kept the same
leads to $\mu\approx 3.4$ implying at least a $3\sigma$ confidence provided that there is no bias due to systematic
errors (see discussion). 

%\noindent
%We note that, using the upper limits  $f_{2,K}$ instead of $f_K$, yields $\gamma\approx 1.33$ and our previous result has vanished, i.e. the %confidence in the result has dropped considerably below $0.675\sigma$ (50\%).
%However, when we insert three times $t_{\rm cr, GC}$ into the equation for $\kappa$, we obtain
%$\kappa=5.13/f_{\rm GC} \approx 1.17$ and the confidence in our previous result drops below 1$\sigma$.

\section{Discussion and conclusions}

%When we assume that GCs are {\bf Roche volume} underfilling at the time of their formation
%they should expand to fill their {\bf Roche volume}s on the two-body relaxation time scale $t_{\rm rx}$
%as has been shown previously by Engle (1999).
%The exact {\bf Roche volume} filling times can be determined with improved precision by parameter studies 
%with ``state-of-the-art'' direct $N$-body simulations. 
%We have $T_{\rm orb} < t_{\rm rx}$ for Milky Way GCs, and the material 
%outside of $r_J$ will continually be removed by the shear forces of the tidal field
%in the course of the orbital evolution such that the {\bf Roche volume} filling condition can be only fullfilled
%in the pericenter of eccentric GC orbits.

The results of the present study are as follows:

\begin{enumerate}
\item We found under the assumptions stated below that GCs are
generally Roche volume underfilling in terms of $\widehat{\lambda}=r_\mathrm{t}/r_\mathrm{J}$.
In the pericenters of their orbits a significant fraction might be Roche volume 
overfilling dependent on the dynamical compression of the outer 
shells compared to the smaller Jacobi radius.
\item We found under the assumptions stated below 
with at least $3\sigma$ confidence that the ratio $\lambda=r_h/r_J$ of half-mass and Jacobi radius
is tendentially larger for an average open cluster within the nearest kpc of the Sun 
than for an average globular cluster and quantified the 
proportionality factor

\be
\mu=\left(\frac{r_h}{r_J}\right)_{\rm OCs}\Big{/}\left(\frac{r_h}{r_J}\right)_{\rm GCs} \approx 3-5.
\ee

\item The difference between OCs and GCs seems to be that, with respect to the concentration, the 
average OC has low concentration 
while the average GC has a high concentration. 
\item  A fraction of OCs may be Roche volume overfilling. However, the simple assumption of virial equilibrium 
breaks down for Roche volume overfilling clusters.
\item A closer inspection of \citet{Baumgardt2010} suggests that there is a physically extended subsample 
of low mass GCs with $r_\mathrm{h} > 10$\,pc, which may represent the post core collapse sequence of 
dissolving clusters. Our sample of GCs with known orbits is too small to confirm this scenario. 
\end{enumerate}

We make the following remarks:

\begin{enumerate}
\item The fact that $\lambda_{\rm GC}\ll 1$ explains (i)
why GCs are spherically shaped as compared to the often irregularly shaped OCs, (ii) why 
GCs are stable against dissolution over a Hubble time and (iii) why
not many GC tidal tails have been found observationally.
\item A fraction of OCs may be Roche volume overfilling ($\lambda_{\rm OC}>1$) at the time of their formation,
but the shear forces of the tidal field will rapidly remove the material outside the Jacobi radius. 
\item Only if the star cluster is in virial equilibrium Eqn. (\ref{eq:kepler}) is valid. Therefore the correlation found by
\citet{vandenBergh1994} suggests that an average GC is in virial
equilibrium if we postulate that a constant ratio $\lambda_{\rm GC}$ 
is reasonable with respect to the general structure of GCs. We emphasize that the virial theorem cannot be 
valid for Roche volume overfilling clusters.
\end{enumerate}

We rely on the following assumptions, stated in order of importance according to our view:

\begin{enumerate}
%\item That the error propagation is quadratic and correlations can be neglected.
%Note that $r_h$ and $\sigma_0$ are related for bound systems, 
%as well as $R$ and $V$.
\item That the orbits of GCs in the sample of \citet{Dinescu1999} can be approximated by
orbits in a purely isothermal halo for which the total angular momentum is conserved.
\item That the orbits of OCs can be approximated by circular orbits \citep[for orbit calculations of OCs
see, e.g.,][]{Cararro1994}.
\item That there are no selection effects concerning the GC sample, i.e. the sample of GCs in \citet{Dinescu1999} is  
representative for the GC population of the Milky Way.
\item That the half-light and projected half-mass radii coincide.
\item That the velocity dispersion of OCs is not too much biased due to the presence of binaries.
%\item That the ``dark matter halo'' in which the GCs orbit are integrated is correctly modelled
%in the JSH95 and P90 models of the Milky Way of Dinescu et al. (1999).
\end{enumerate}

It is crucial to this investigation whether the approximations (i) and (ii) are justified.
In the future, our results may be falsified or improved towards
higher confidence levels when more and better data are available.

Our investiation suggests that most GCs were formed deep in their potential well, i.e. Roche volume 
underfilling in contrast to OCs, which can be even Roche volume overfilling after gas expulsion. 
In future work we plan to investigate the dynamical reasoning for 
these intrinsic differences and to quantify the impact on the 
dissolution process of the star clusters.

\section{Acknowledgements}

AE is grateful for support by grant JU 404/3-1 of the German Research Foundation (DFG) 
and thanks Prof. Dr. Eva Grebel for a comment made during a talk which directed his interest
to the van-den-Bergh correlation.
Discussions with Prof. Dr. Rainer Spurzem are gratefully acknowledged and the fact, that
the idea to Eqn. (\ref{eq:ratio1}) was first written down on a notepad of him.
Both authors thank the referee for the thoughtful comments.

\appendix

\section{The ratio $\beta_C$}

\label{sec:betac}

%\begin{figure}
%\includegraphics[width=0.5\textwidth]{hillsurf.pdf} 
%\caption{The Roche or Hill surface.} 
%\label{fig:hillsurf}
%\end{figure}

%Figure \ref{fig:hillsurf} shows the Roche or Hill surface which contains the Roche volume in its interior.
%In the $x$ direction, its extent is twice the Jacobi radius $r_J$.

For any galactic potential $\Phi$, the dimensionless ratio $\beta_C=\kappa_C/\Omega_C$ is given by

\be
\beta_C^2 = 2\left( 1+\frac{d\ln V_C}{d\ln R} \right)\Big\vert_{R_C} = 3 + R\frac{\left(d^2\Phi/dR^2\right)}{\left(d\Phi/dR\right)}\Big\vert_{R_C}
\ee

\noindent
where $\kappa_C$ and $\Omega_C$ are the epicyclic and circular frequency related to a circular orbit,
$R_C$ is its radius and $V_C=\Omega_C R_C$ is the circular velocity at that radius.
Figure \ref{fig:betaall} shows that $\beta_C$ is approximately constant for a wide range of
Galactocentric radii for three different analytic Milky Way potentials, among them
that two models used in \citet{Dinescu1999}. 

\section{Eccentric orbits}

\label{sec:appecc}

\subsection{Kepler case}

If we approximate the Milky Way potential
by a Kepler potential $\Phi_K\propto r^{-1}$
we find $\beta_{C,K}=1$. Apo- and pericenter are defined by
$R_P = a(1-e)$ and $R_A = a(1+e)$ where $a$ and $e$ are the semimajor axis and the eccentricity of the orbit.
Also, we have the relation (c.f. King 1962, BT2008)

\be
L^2 = \Omega^2 R^4 = G M_g a(1-e^2) =  \frac{GM_g}{a}R_A R_P, \ \ \ 
\ee

\noindent
where $M_g, \Omega$ and $R$ are the mass of the point-like galaxy,
 the angular speed and the galactocentric 
radius at any orbital phase, respectively. 

With the 
gravitational potential and its derivatives

\be
\Phi = -\frac{GM_g}{R}, \ \ \ \frac{d\Phi}{dR} = \frac{GM_g}{R^2}, \ \ \ \frac{d^2\Phi}{dR^2} = -\frac{2GM_g}{R^3}
\ee

\noindent
we obtain the Jacobi radius

\bea
r_J &=& \left[ \frac{G M_{cl}}{\Omega^2 - \frac{d^2\Phi}{dR^2}} \right]^{1/3} \label{eq:genkep} \\
&=& \left( \frac{M_{\rm cl}}{M_g}\right)^{1/3} \left( \frac{aR^4}{R_A R_P + 2 a R} \right)^{1/3}
 \eea

\noindent
for circular and eccentric orbits (cf. King 1962).

%We find

%\be
%\frac{r_{J,A}}{r_{J,P}} = \left(\frac{\Omega_P}{\Omega_A}\right)^{2/3}= \left( \frac{1+e}{1-e} \right)^{4/3} = \left(\frac{R_A}{R_P}
%\right)^{4/3}.
% \ee

%\noindent
%For a circular orbit with $e=0$ we get $r_{J,A}/r_{J,P} = 1$. For $e=0.622 \pm 0.03$ (the median eccentricity of the sample in
%Dinescu et al. 1999) we obtain  $r_{J,A}/r_{J,P} \approx 6.97$. However, this is an extreme case.

\subsection{Isothermal case}

\label{sec:iso}

If we approximate the Milky Way potential
by the potential of an isothermal sphere $\Phi_I= V_C^2 \ln(R/R_0)$ with the circular velocity $V_C$
we find $\beta_{C,I}=\sqrt{2}$ (note that the isothermal sphere has a constant rotation curve). 
%and the equality (\ref{eq:ratio1}) for circular orbits becomes 

% \be
 %\left(\frac{r_h}{r_J} \right)^3 = \left(\frac{t_{\rm orb}}{T_{\rm orb}}\right)^{2}. \label{eq:ratio3} 
 %\ee
 
 \noindent

The energy (say, of a globular cluster) in an isothermal sphere is given by

\be
E = \frac{V^2}{2} + V_C^2\ln\left(\frac{R}{R_0}\right)
\ee

\noindent
where $R_0$ is a lenght unit. This yields

\bea
V^2 &=& 2\left[E - V_C^2\ln\left(\frac{R}{R_0}\right)\right] \nonumber \\
&=& 2V_C^2\ln\left[\exp(E/V_C^2)\left(\frac{R_0}{R}\right)\right] = 2V_C^2 \ln\left(\frac{R_0'}{R}\right)
\eea

\noindent
with $R_0' =\exp(E/V_C^2) R_0$. For simplicity we substitute in the following discussion $R_0' \rightarrow R_0$.

From the angular momentum conservation we obtain at apo- and pericentre

\be
R_P^2\ln(R_0/R_P) = R_A^2\ln(R_0/R_A). \label{eq:1}
\ee

\noindent
From the energy conservation we obtain

\be
\frac{V_P^2}{2} + V_C^2\ln\left(\frac{R_P}{R_0}\right) = \frac{V_A^2}{2} + V_C^2\ln\left(\frac{R_A}{R_0}\right)\label{eq:2}
\ee

%\noindent
%Eqn. (\ref{eq:1}) yields

%\be
%\ln(R_0) = \frac{R_A^2\ln(R_A)-R_P^2\ln(R_P)}{R_A^2-R_P^2}\label{eq:3}
%\ee

%\noindent
%Inserting Eqn. (\ref{eq:3}) in (\ref{eq:2}) yields after some calculations
\noindent
We also obtain 

\be
V_P^2 - V_A^2 %= \Omega_P^2 R_P^2 -\ \Omega_A^2 R_A^2 
= 2 V_C^2 \ln(R_A/R_P)
\ee

%\noindent
%This yields

%\be
%\frac{r_{J,A}}{r_{J,P}} = \left(\frac{\Omega_P}{\Omega_A}\right)^{2/3}= \left[ \frac{\ln(R_P)}{\ln(R_A)} \right]^{1/3} \left(\frac{R_A}{R_P}
%\right)^{2/3}.
%\ee

\noindent
The angular momentum as a function of $R_A$ and $R_P$ is given by

\be
L = V_C R_P R_A \sqrt{\frac{2\ln(R_A/R_P)}{R_A^2-R_P^2}} \label{eq:angmom}
\ee

\noindent
The limiting cases are circular and radial orbits. Using $L^2$ we have verified (1) with the rule of l'Hospital that
the Eqn. (\ref{eq:angmom}) is consistent with the limiting case of a circular orbit with radius $R_C$ and velocity $V_C$ 
and (2) that the radial orbit has zero angular momentum. From Eqn. (\ref{eq:angmom}) follows the relation

\be
L^2 = \Omega^2 R^4 =  2 V_C^2 R_P^2 R_A^2 \frac{\ln(R_A/R_P)}{R_A^2-R_P^2}
\ee

\noindent
where $\Omega$ is the angular velocity at any orbital radius $R$. We follow now the derivation of $r_J$ in \citet{King1962}
for the case of an isothermal sphere. The gravitational potential of the isothermal sphere and its derivatives are given by

\be
\Phi = V_C^2\ln\left(\frac{R}{R_0}\right), \ \ \ \ \ \frac{d\Phi}{dR}= \frac{V_C^2}{R}, \ \ \ \ \ \frac{d^2\Phi}{dR^2} = -\frac{V_C^2}{R^2}.
\ee

\noindent
We obtain for the Jacobi radius

\bea
r_J^3 &=& \frac{G M_{cl}}{\Omega^2 - \frac{d^2\Phi}{dR^2}} \label{eq:geniso} \\
&=& \frac{G M_{cl} (R_A^2 - R_P^2) R^4}{2V_C^2R_P^2 R_A^2 \ln(R_A/R_P) + V_C^2 (R_A^2 - R_P^2) R^2} \label{eq:gen}
\eea

\noindent
for circular and eccentric orbits (cf. King 1962).

If we define a ``guiding radius''

\be
R_g  = \frac{L}{V_C}  \label{eq:guiding}
\ee

\noindent
the Jacobi radius can be written as

\be
r_J = r_v^{1/3} \left(\frac{R^4}{R_g^2 + R^2} \right)^{1/3} 
\ee

\noindent
where $r_v = GM_{cl}/V_{\rm C}^2$ is the velocity radius given by Eqn. (\ref{eq:rv}) with the constant $V_C$ of the isothermal sphere. 
We also obtain

\be
\lambda_{\rm I} = \left(\frac{r_h}{r_J}\right)_{\rm I} = \left(\frac{t_{\rm orb}}{2\pi}\right)^{2/3} \left(\frac{V_C}{\sqrt{2}R(t)}\right)^{2/3} \left( 1 + \frac{R_g^2}{R(t)^2}\right)^{1/3}.
\ee

\noindent
where $R(t)$ is time-dependent for an accentric orbit.

%If we define similarly to the Kepler case $R_A = R_m(1+e)$ and $R_P = R_m(1-e)$, where $R_m$ and $e$ are 
%``mean radius'' and ``eccentricity'' the guiding radius can be written as 

%\be
%R_g^2 = R_m^2\frac{(1-e^2)^2}{2e}\ln\left(\frac{1+e}{1-e}\right). \label{eq:guiding2}
%\ee

%\noindent
%and

%\be
%\lim_{R\rightarrow\infty} r_J \propto R^{2/3}.
%\ee

%\noindent
%The limiting cases are circular orbits ($R_P \rightarrow R_A$) and radial orbits ($R_P \rightarrow 0$).
%For the limiting cases the rule of l' Hospital must be used. 

%In these cases we obtain exact power laws since the leading constant in the denominator of Eqn. (\ref{eq:gen}) vanishes:

%For the circular orbits we have

%\be
%\lim_{R_P\rightarrow R_A} r_J^3 = \frac{GM_{\rm cl}(R_A^2 - R_A^2) R^4}{2V_C^2 R_A^4 \ln(1) + V_C^2(R_A^2 - R_A^2)R^2} \propto R^2
%\ee

%For the radial orbits we have
%\be
%\lim_{R_P\rightarrow 0} r_J^3 = \frac{GM_{\rm cl}R_A^2 R^4}{2V_C^2 0^2 R_A^2 \ln(R_A/0) + V_C^2 R_A^2 R^2} \propto R^2
%\ee

%\noindent
%The general case Eqn. (\ref{eq:gen}) lies between these two extrema.

\subsection{Harmonic case}

In the case of a sphere with homogeneous density $\rho_0$ we have  $\beta_{\rm C,H} = 2$ and
$\Omega= \sqrt{4\pi G\rho_0/3}$ which is independent of $R$. 
Both the Jacobi radius in Eqn. (\ref{eq:rj1}) and the ratio in Eqn. (\ref{eq:ratio1}) are not well defined.

\subsection{Beyond the solar circle}

\begin{figure}
\includegraphics[width=0.5\textwidth]{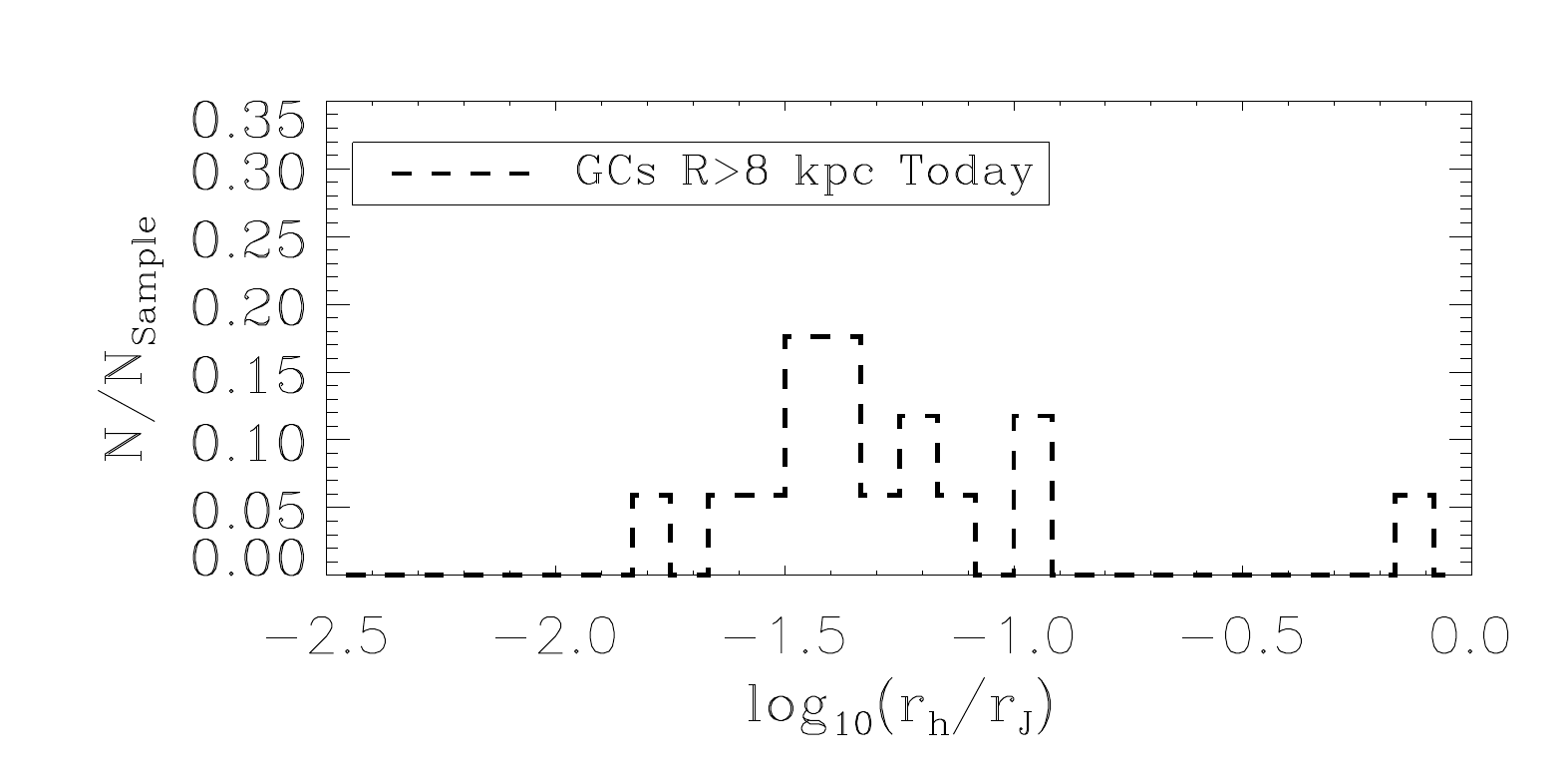} 
\caption{The ratio $r_{h, \rm 3D}/r_J$ of GCs today  calculated from Eqn. (\ref{eq:ratio2})
 for 17 GCs out of the 34 GCs in the sample by \citet{Dinescu1999} for which $R>8$ kpc
under the assumption that the GCs are moving in an isothermal halo with $V_C\approx 228.5$ km/s and that the van-den-Bergh correlation holds.} 
\label{fig:histsgcs58}
\end{figure}

\citet{Baumgardt2010} show that clusters with galactocentric distances $R > 8$ kpc fall into two distinct groups: one group 
of compact, tidally-underfilling clusters with $r_h/r_J <$ 0.05 and another group of tidally filling clusters which 
have $0.1 < r_h/r_J < 0.3$. In Figure \ref{fig:histsgcs58} we calculated the ratio $r_{h, \rm 3D}/r_J$ of GCs today 
from Eqn. (\ref{eq:ratio2}) for 17 GCs out of the 34 GCs in the sample by \citet{Dinescu1999} for which $R>8$ kpc
under the assumption that the GCs are moving in an isothermal halo with $V_C\approx 228.5$ km/s as in Figure \ref{fig:histsgcs3}. 
We do not clearly see the dichotomy in Figure \ref{fig:histsgcs58}. The reason may be that the \citet{Dinescu1999} 
data set is not large enough.

However, \citet{Baumgardt2010}  argued about the case of NGC 2419, where the best-fitting King model has a tidal radius
of 150 pc, while the estimated Jacobi radius of the cluster is around 800 pc (derived in Baumgardt et al. 2009). 
From our point of view this is a perfect case of a GC embedded deeply in its own potential well, such that a King 
model of an isolated cluster is a very good approximation, because the tidal field is negligible for the cluster.

\label{lastpage}


\begin{thebibliography}{99}
\bibitem[\protect\citeauthoryear{Baumgardt et al.}{2009}]{Baumgardt2009}  Baumgardt H., et al., 2009, MNRAS, 396, 2051
\bibitem[\protect\citeauthoryear{Baumgardt et al.}{2010}]{Baumgardt2010}  Baumgardt H., Parmentier, G., Gieles, M., Vesperini, E., 2010, MNRAS, 401,1832
\bibitem[\protect\citeauthoryear{Binney \& Tremaine}{2008}]{Binney2008} Binney J., Tremaine, S., 2008, {\it Galactic dynamics}, $2$nd ed., Princeton University Press, Princeton
\bibitem[\protect\citeauthoryear{Bonatto \& Bica}{2011}]{Bonatto2011} Bonatto, C., Bica, E., 2011, MNRAS, 415, 2827
\bibitem[\protect\citeauthoryear{Cararro \& Chiosi}{1994}]{Cararro1994} Carraro, G., Chiosi, C., 1994, A\&A, 288, 751
\bibitem[\protect\citeauthoryear{Converse \& Stahler}{2010}]{Converse2010} Converse, J. M., Stahler, S. W., 2010, MNRAS 405, 666
\bibitem[\protect\citeauthoryear{Dinescu et al.}{1999}]{Dinescu1999} Dinescu D. I., Girard T. M., Altena W. F., 1999, AJ, 117, 1792
\bibitem[\protect\citeauthoryear{Ernst et al.}{2010}]{Ernst2010} Ernst A., Just, A., Berczik, P., Petrov, M.I., 2010, A\&A 524, A62
\bibitem[\protect\citeauthoryear{Guerkan et al.}{2004}]{Guerkan2004} G\"urkan, A., Freitag, M., Rasio, F. A., 2004, Ap. J., 604, 632
\bibitem[\protect\citeauthoryear{Harris}{1996, 2010 edition}]{Harris2010} Harris, W.E. 2012, arXiv:1012.3224
\bibitem[\protect\citeauthoryear{Hong et al.}{2004}]{Hong2004} Hong, J., Schlegel, E. M., Grindlay, J.E., 2004, Ap. J., 614, 508 
\bibitem[\protect\citeauthoryear{Johnston et al.}{1995}]{Johnston1995} Johnston, K. V., Spergel, D. N., Hernquist, L., 1995, Ap. J., 451, 598 (JSH95)
\bibitem[\protect\citeauthoryear{Just et al.}{2009}]{Just2009} Just, A., Berczik, P., Petrov, M.I., Ernst, A., 2009, MNRAS 392, 969
\bibitem[\protect\citeauthoryear{Kharchenko et al.}{2009}]{Kharchenko2009} Kharchenko N. V., Berczik P., Petrov M. I., Piskunov A. E., R{\"o}ser S., Schilbach E., Scholz R.-D. 2009, A\&A, 495, 807
\bibitem[\protect\citeauthoryear{King}{1962}]{King1962} King I. R., 1962, AJ, 67, 471
\bibitem[\protect\citeauthoryear{Kroupa}{2001}]{Kroupa2001} Kroupa, P., 2001, MNRAS, 322, 231
\bibitem[\protect\citeauthoryear{K\"upper et al.}{2008}]{Kuepper2008} K\"upper A. H. W., Macleod A., Heggie D. C., 2008, MNRAS, 387, 1248
\bibitem[\protect\citeauthoryear{Lamers \& Gieles}{2006}]{Lamers2006} Lamers, H. J. G. L. M.,Gieles, M., 2006, A\&A 455, L17
\bibitem[\protect\citeauthoryear{Ortolani et al.}{2012}]{Ortolani2012} Ortolani, S., Bonatto, C., Bica, E., Barbuy, B., Saito, R. K., 2012, AJ, 144, 147
 \bibitem[\protect\citeauthoryear{Paczy{\'n}ski}{2011}]{P90} Paczy{\'n}ski, B., 1990, Ap. J., 348, 485 (P90)
\bibitem[\protect\citeauthoryear{Piskunov et al.}{2006}]{Piskunov2006} Piskunov, A. E., Kharchenko, N. V., R\"oser, S., Schilbach, E., Scholz, R.-D., 2006, A\&A, 445, 545
\bibitem[\protect\citeauthoryear{Piskunov et al.}{2007}]{Piskunov2007} Piskunov, A. E., Schilbach, E., Kharchenko, N. V., R\"oser, S.,  Scholz, R.-D., 2007, A\&A, 468, 151
\bibitem[\protect\citeauthoryear{van den Bergh}{1994}]{vandenBergh1994} van den Bergh, S., 1994, AJ, 108, 2145
\end{thebibliography}
\end{document}